\input harvmac
\overfullrule=0pt

\lref\sknd{H.\ J.\ Boonstra, B.\ Peeters and K.\ Skenderis, Nucl.\ Phys.\
{\bf B533} (1998) 127-162, hep-th/9803231.}
\lref\mms{E. Martinec, G. Moore and N. Seiberg, Phys. Lett. {\bf 263B}
(1991) 190.}
\lref\jcka{E. D'Hoker, D. Freedman and R. Jackiw, Phys. Rev. {\bf D28}
(1983) 2583.} 
\lref\ban{ M. Banados, K. Bautier, O. Coussaert, M. Henneaux and M.
     Ortiz, hep-th/9805165.}
\lref\tlee{T. Lee, hep-th/9806113.}
\lref\emads{E. Martinec, hep-th/9804111.}
\lref\bbg{ K. Behrndt, I. Brunner and I. Gaida, hep-th/9806195.}
\lref\gks{ A. Giveon, D. Kutasov and N. Seiberg, hep-th/9806194.}
\lref\twodg{V. Knizhnik, A. Polyakov and A. Zamolodchikov, Mod. Phys. Lett.
{\bf A3} (1988) 819.} 
\lref\twoda{F. David, Mod. Phys. Lett. {\bf A3} (1988) 1651.}
\lref\twodb{ F. David, J. Distler and H. Kawai, 
Nucl. Phys. {\bf B321} (1989) 509.}
\lref\stp{A. Sevrin, W. Troost and A. Van Proeyen, Phys. Lett {\bf B208}
(1988) 447.}
\lref\wtop{E. Witten, Commun.Math.Phys.118 (1988) 411; Phys. Lett. {\bf 206B} (1988) 601. }
\lref\jmtwo{N. Itzhaki, J. Maldacena, J. Sonnenschein and S.
Yankielowicz, hep-th/9802042.}
\lref\dbo{J. deBoer, hep-th/9806104.}
\lref\sez{S. Deger, A. Kaya, E.Sezgin and P. Sundell,
hep-th/9804166.}
\lref\qtdg{S. B. Giddings and A. Strominger, Phys. Rev. {\bf D47} (1993) 2454.}
\lref\bfss{T. Banks, W. Fischler, S. Shenker and L. Susskind, 
Phys. Rev. {\bf D55} (1997) 5112.}
\lref\dlcq{L. Susskind, hep-th/9704080.}
\lref\romans{L. Romans, Nucl. Phys. {\bf B276} (1986) 71.} 
\lref\kimn{H. Kim, L. Romans and P. Nieuwenhuizen, Phys. Rev. {\bf D32} (1985)
389.}
\lref\vw{C. Vafa and E. Witten,
 Nucl. Phys. {\bf B431} (1994) 3,
 hep-th/9408074.}
\lref\sei{N. Seiberg, Prog. Th. Phys.{\bf 102} (1990) 319.}
\lref\jck{E. D'Hoker and R. Jackiw, Phys. Rev. Lett. {\bf 50} 
(1983) 1719; Phys. Rev. {\bf D26} (1982) 3517.}
\lref\ascv{A.\ Strominger and C.\ Vafa,
 { Phys. Lett.} {\bf B379}
(1996) 99, hep-th/9601029.}
\lref\cl{M. Cvetic and F. Larsen, hep-th/9805097; hep-th/9805146.}
\lref\nhbh{A. Strominger, hep-th/9712251.}
\lref\ho{G. Horowitz and H.Ooguri, hep-th/9802116.}
\lref\dps{M.\ Douglas, J.\
Polchinski and A.\ Strominger, hep-th/9703031.}
\lref\jm{J.\ Maldacena,
hep-th/9711200.}
\lref\msads{J.\ Maldacena and A.\ Strominger,
hep-th/9804085.}
\lref\hyun{S.\ Hyun, hep-th/9704005.}
\lref\bph{H.\ Boonstra, B.\ Peeters, and K.\ Skenderis, hep-th/9706192.}
\lref\ss{K.\ Sfetsos and K.\ Skenderis, hep-th/9711138.}
\lref\btz{M.\ Banados, C.\ Teitelboim and J.\ Zanelli, {\sl
Phys. Rev. Lett.} {\bf 69} (1992) 1849.}
\lref\bhtz{M.\ Banados, M.\ Henneaux, C.\ Teitelboim, and J.\ Zanelli, 
{Phys.Rev.} {\bf D48} (1993) 1506, gr-qc/9302012.}
\lref\dlas{D. A . 
Lowe and A. Strominger, Phys. Rev. Lett {\bf 73} (1994) 1468.}
\lref\jc{J. A. Cardy, Nucl. Phys. {\bf B270} (1986) 186.}
\lref\kal{P. Claus, R. Kallosh, J. Kumar, P. Townsend and A. van Proeyen,
hep-th/9801206.}
\lref\nhg{M. Cvetic and A. Tseytlin, Phys. Rev. {\bf D53} (1996) 5619.}
\lref\jbmh{J.\ D.\ Brown and M.\
Henneaux, { Comm. Math. Phys.} {\bf 104} (1986) 207.}
\lref\hms{G. Horowitz, J. Maldacena and A. Strominger,
Phys. Lett. {\bf B383} (1996) 151, hep-th/9603109.  }
\lref\withol{E. Witten,
 hep-th/9802150.}
\lref\msu{
J. Maldacena and L. Susskind, 
Nucl.Phys. {\bf B475} (1996) 679, hep-th/9604042.}
\lref\gkp{S.S. Gubser, I.R. Klebanov and A.M. Polyakov,
hep-th/9802109.}
\lref\jmas{J. Maldacena and A. Strominger, Phys.Rev. {\bf D55} (1997) 861.} 
\lref\skend{ H. J. Boonstra, B. Peeters and  K. Skenderis,
hep-th/9803231. }

\def\ads{$AdS_3$}
\def\ad{$AdS_2$}
\def\slr{{SL(2,R)}}

\def\p{\partial}

\Title{hep-th/9809027}{\ad\ Quantum Gravity and String Theory}

\centerline{Andrew  Strominger}
\bigskip\centerline{Department of Physics}
\centerline{Harvard University}
\centerline{Cambridge, MA 02138} 

\vskip .3in
\centerline{\bf Abstract}
\smallskip
\ad\ has an $\slr$ isometry group. It is argued that 
in the context of quantum gravity on \ad\ this group is enlarged 
to the full infinite-dimensional $1+1$ conformal group.
Massive scalar fields are coupled to \ad\ gravity and shown to have 
associated conformal weights $h(m)$ shifted by their mass.
For integral values of 
$h$ primary boundary operators are obtained as $h$ normal derivatives of 
the scalar field. 
\ad\ string theories arise in the 
`very-near-horizon' limit of $S^1$-compactified \ads\ string 
theories. This limit corresponds to energies far below 
the compactification scale.
The dual conformal field theory has one null dimension and can 
in certain cases be described as the discrete light cone quantization 
of a two-dimensional deformed symmetric-product conformal 
field theory. Evidence is given that the 
\ad\ Virasoro algebra is related to the 
right-moving \ads\ Virasoro algebra by a topological twist which shifts 
the central charge to zero.

\Date{}
\newsec{Introduction}
Perhaps the best - understood of the family of $AdS$/CFT 
dualities \jm\ is the case \ads /2D-CFT \refs{\jbmh, \jm, \nhbh \msads \sez
\dbo \gks 
\bbg \emads \cl \tlee - \ban},
largely because the conformal group is infinite dimensional and 
greatly constrains the dynamics. 
In higher dimensions the field theory side is less well-understood, 
and the duality has largely been used to learn about 
field theory from gravity. In lower dimensions - namely the \ad\ case - 
one confronts the puzzles of black holes undistracted by inessential 
complications. Here one hopes to learn about gravity from field 
theory, and new wrinkles are encountered. One such wrinkle
 is that \ad\ has two 
timelike boundaries, one originating from the region inside and one 
from the region outside the horizon. 
Hence the dual one-dimensional conformal quantum mechanics does not 
appear to live on a connected manifold.

At present very little is understood about this \ad\ case, which is 
the subject of this paper. Some previous investigations appear in 
\sknd. 
We first consider general properties of two-dimensional quantum 
gravity plus matter on \ad. While the subject of two-dimensional 
quantum gravity 
has received considerable attention\foot{A review with 
emphasis relevant to the 
present context can be found in the first few sections of reference \sei.}, 
we shall see that \ad\ 
provides an interesting new context. We then consider properties and dualities 
of \ad\ string theories obtained by compactification of the 
well - understood \ads\ case. 

All theories of quantum gravity in two 
dimensions are conformal field theories \refs{\twodg \twoda -\twodb}. 
Liouville-like theories with \ad\ ground states -- 
which necessarily are open-string-like theories on the 
strip -- are no exception.  A key distinction of 
the \ad\ case discovered in \refs{\jck ,\jcka} is that, unlike 
on the circle or the plane,  the theory on the strip has a ground state 
which is annihilated by the global $\slr$ subgroup of the conformal group. 
This subgroup is identified with the \ad\ isometry group. We 
accordingly argue in section 2.1 that the global $\slr$ conformal
symmetry in the $AdS_2/CFT_1$ duality is enlarged to the 
full local conformal group, much as in the $AdS_3/CFT_2$ case \jbmh . 
In section 2.2 we consider a scalar field $\phi$ of mass $m$ 
coupled to gravity on \ad. It is shown perturbatively 
that the effect of the mass term is to 
shift the dimension of the highest weight state created by 
the scalar field to $h=\half(1+\sqrt{1+m^2\ell^2})$, where $\ell$ 
is the \ad\ radius. The corresponding 
boundary operator for integral $h$ is $(\p_\sigma)^h \phi$, 
with $\p_\sigma$ the normal derivative. This is primary because 
lower derivatives of $\phi$ vanish 
on the boundary. Section 2.3 contains a discussion of deficit angles 
produced by boundary insertions and a consequent upper bound on 
allowed masses. 

In section 3 we turn to string theory. The near-horizon geometry of 
the five-dimensional black hole of \ascv\ is \ads\ 
periodically identified on a circle of 
asymptotic radius $R$ \hyun. In section 2.1 the interpretation of 
this as a compactification to two dimensions is presented. 
In section 3.2 it is shown that the geometry of the `very-near-horizon' 
limit, corresponding to energies much less than $ 1 \over R$, 
is \ad\ with a constant 
Kaluza-Klein $U(1)$ field strength. It is also found that in this limit 
the \ads\ identification lies entirely within the $\slr_L$ subgroup of 
the $\slr_L \times\slr_R$ isometry group. The dual description of the 
very-near-horizon 
limit is argued in section 3.3 to be the DLCQ of the deformed 
symmetric product 
field theory described in \ascv. Hence in this context the 
one dimension of the conformal field theory in  the \ad 
$\leftrightarrow CFT_1$ duality 
is null. This is naturally interpreted as half of a 2D 
conformal field theory rather than as conformal quantum mechanics. 
In section 3.4 the two-dimensional effective action arising 
from compactification of 
\ads\ is described.

Symmetries are discussed in section 4. 
The \ads\ string theory admits the action of two Virasoro 
algebras with large central 
charges. Only the right action survives the 
compactification. \ad\ on the other hand 
admits the action of a single Virasoro algebra which has $c=0$ 
since it is a theory of gravity.
In section 4.1 we argue, assuming the existence of a suitable 
Hilbert space on \ad,  that the 
\ad\ Virasoro algebra is related to the \ads\ algebra by 
a twisting which shifts the central charge to zero. This 
twisting is similar or identical 
to that used in earlier constructions of topological field theories and 
topological gravity 
from ordinary supersymmetric conformal field theories \refs{\wtop}. In 
section 4.2 the analysis 
of neutral scalar fields of section 2.2 is extended to the $U(1)$ charged
scalar fields arising in \ads\ compactification.  It is found that due to 
cancellations the  lowest
conformal weight of the states created by such fields is independent of the 
charge and the Kaluza-Klein mass correction, and hence can be small 
even for very 
massive fields. 

It should be noted 
\ad\ string theories arise as the near-horizon limits of a wide variety
of four and five -dimensional black holes without passing through an 
intermediate \ads\ region. It is not clear to us which of the features 
discovered in sections 3 and 4 are generic to all \ad\ string theories. 
The discussion of section 2 on the other hand, and in particular the 
enlargement of the $\slr$ to the full conformal group, pertains to 
all theories of quantum gravity on \ad.

\newsec{\ad\ Quantum Gravity as Conformal Field Theory}
In this section we consider general properties of 
two-dimensional quantum gravity coupled to matter with an \ad\  
ground state. We take 
Newton's constant to be
small and the \ad\ radius large 
so that a semiclassical description is appropriate. 
Since \ad\ has two timelike boundaries, such theories 
live on the strip.\foot{We will mainly consider the covering space 
of \ad, which for brevity will be referred to simply as \ad.} 
String theory on \ad $\times M^8$  can be viewed as two-dimensional
gravity with an infinite tower of fields arising from massive 
string and Kaluza-Klein modes. In conformal 
gauge 
\eqn\cgf{ds^2=-e^{2\rho}dt^+dt^-,} 
every two-dimensional quantum theory of gravity is a $c=0$ (including 
ghosts) 
conformal field theory, with 
conformal invariance arising as a result of  diffeomorphism 
invariance and background independence. Hence \ad\ string theory 
can be viewed as conformal field theory on the strip.
This is a lower dimensional analog 
of the philosophy of \refs{\jbmh -\nhbh} in which \ads\ 
string theory is viewed as conformal field theory on the 
circle.
Since the conformal transformations are residual
diffeomorphisms, not all of the states in the theory should be regarded 
as inequivalent physical states.  In a BRST formalism physical 
states will be identified as usual as BRST cohomology classes. 
In a ghost-free gauge one has a physical state condition of 
the form $L_n|\psi>=0$. These are the 
quantum versions of the constraints which determine the metric 
in terms of the matter fields.

\subsec{ \ad\ Isometries and the Virasoro Algebra}
The gravitational degrees 
of freedom of \ad\ quantum gravity 
in conformal gauge are described 
by a Liouville theory governing the metric conformal factor
$\rho$, possibly generalized by dilaton couplings. The action 
for a specific case in string theory is 
given in section 3 below.   
\ad\ quantum gravity is like an 
open string theory on the strip owing to the two boundaries 
of \ad. Hence there is only one copy of the Virasoro algebra. 
This has an $\slr$ subgroup which generates the 
usual transformations
\eqn\fgep{\eqalign{L_{-1}&={i \over 2}(e^{-2iu^+}\p_++e^{-2iu^-}\p_-) ,\cr
 L_{0}&= {i \over 2}(\p_++\p_-),\cr
  L_{1}&= {i \over 2}(e^{2iu^+}\p_++e^{2iu^-}\p_-) ,\cr }}
where $u^\pm=\half (\tau \pm \sigma )$ and $0\le \sigma \le \pi$. 
As noted some time ago by D'Hoker 
and Jackiw \refs{\jck ,\jcka},\foot{The theory
  considered in \refs{\jck ,\jcka} differs in the scalar couplings 
from the string theory case considered below but 
this is not relevant for the following.}  the Liouville 
theory with cosmological constant $-{2 \over \ell^2}$ 
has an $\slr$-invariant vacuum characterized by 
\eqn\dfh{\ell e^{-\rho}=sin(u^+-u^-).}
This vacuum is of course the \ad\ spacetime. 
The  $\slr$ generators \fgep, originally introduced without 
reference to a metric, are 
the $\slr$  Killing vectors of \ad. 

This situation contrasts with Liouville 
theory on the circle or plane, for which there is no 
vacuum preserving the $\slr_R \times \slr_L$ Virasoro subgroup 
except for the singular configuration $e^\rho=0$. The absence of such 
an invariant vacuum has greatly complicated the analysis of 
the theory on the circle.  Hence in 
this sense Liouville theory lives more naturally on the strip than 
on the plane. 

An important conclusion here is that the global $\slr$ 
symmetry of \ad\ string theory is enhanced to (half of) 
the $1+1$ conformal group, much as the global 
$\slr_L \times \slr_R$ symmetry of \ads\ string theory is 
enhanced to the full $1+1$ conformal group. This conclusion 
follows from general properties of \ad\ quantum gravity, and so 
applies to all \ad\ string theories.

\subsec{Coupling to Massive Scalars}
Two-dimensional gravity may also contain scalar fields 
with actions of the form
\eqn\scac{S_\phi=-{1 \over 4 \pi}\int d^2x\sqrt{-g}\bigl( (\nabla \phi)^2
+m^2\phi^2)\bigr)={1 \over 8 \pi}\int d^2u\bigl( 4\p_+ \phi\p_-\phi
-{m^2 } e^{2\rho}\phi^2)\bigr) .}
We wish to study the dynamics of such scalar fields in a perturbation expansion
about the \ad\ vacuum \dfh.
From the kinetic term we see that $\phi$ has global scaling dimension 
zero. The mass term is nevertheless a marginal perturbation because 
$e^{2\rho}$ is dimension $(1,1)$ operator.\foot{An explicit demonstration 
in a context close to the present one can be found in \qtdg.}  
The wave equation 
\eqn\sdf{\nabla^2 \phi = m^2\phi,}
reduces in conformal gauge to 
\eqn\cfdh{-4e^{-2\rho}\p_+\p_- \phi=m^2\phi.}
It is easily verified the \ad\ laplacian defined by the metric 
\dfh\ is precisely $-{4 \over \ell^2}$ times the $\slr$ Casimir obtained by squaring \fgep. 
Hence 
\eqn\fkl{-4(-L_0(L_0-1)+L_{-1}L_1)\phi=m^2\ell^2\phi .}
For primary, normalizable  solutions $\phi_h$ of \fkl\ with $L_1\phi_h=0$ and 
$L_0=h$ 
\eqn\ghj{4h(h-1)\phi_h=m^2\ell^2\phi_h}
or 
\eqn\hmn{h={1 \over 2}(1+\sqrt{1+m^2\ell^2}),}
in agreement with \refs{\gkp,\withol,\msads}.
$L_1\phi_h=0$ then implies 
\eqn\phj{\phi_h=(e^{-2iu^+}-e^{-2iu^-})^h.}

Further solutions of \fkl\ are generated by acting 
on $\phi_h$ with $L_{-1}$, as this commutes with the $\slr$ 
Casimir operator: 
\eqn\bdf{B_n(\sigma, \tau)=b_n(\sigma )e^{-i(n+h) \tau}=
C_n (L_{-1})^n \phi_h ,~~~~n \ge 0.}
Together with their complex conjugates $B^*_n$ these are a complete 
set of modes. We choose the  normalization constants $C_n$ so that  
\eqn\fth{\langle B_m | B_n \rangle = \delta_{mn}}
with respect to the Klein-Gordon inner product 
\eqn\kgi{\langle A| B \rangle =-i\int_0^\pi 
d \sigma (A^*\p_\tau B-\p_\tau A^* B).} 
$\phi$ may then be promoted to a quantum field and expanded in an
oscillator basis
\eqn\phg{\phi=\sum_{n=0}^\infty a^\dagger_n B_n^*+a_nB_n.}
The canonical commutation relations for $\phi$ then imply as usual 
\eqn\dalp{[a_m, a^\dagger_n]=\delta_{mn}.}
Using the $\phi$ stress tensor derived from the action \scac, 
one finds that the quantum Virasoro generator $L^\phi_0$ gets a contribution
\eqn\lmn{L^\phi_0=\sum_{n=0}^\infty (h+n)a^\dagger_na_n.}
It follows that, as expected, $a^\dagger_n$ creates a state 
with $L^\phi_0=h+n$. It is also possible to see that  $[L_1^\phi ,a_0^\dagger]=0$
so that $a_0^\dagger|0\rangle $, where $|0\rangle $ is defined by 
$a_n |0\rangle=0 $, is a highest weight state.\foot{The  Virasoro constraints 
dictate the gravitational dressing of this state, which is higher order in 
the perturbative regime of small Newton's constant considered here.}

Therefore for every field of mass $m$ there is a highest weight state
of weight \hmn\ in the conformal field theory. 
In conformal field theory on the strip there is a one-to-one 
correspondence\foot{Modulo possibly relevant subtleties discussed in \sei.} between highest weight states 
and primary operators at the boundary. For integral 
$h$ the primary operator 
corresponding to $a_0^\dagger|0\rangle $ is 
\eqn\popp{(\p_\sigma)^h \phi(0,\tau).}
Ordinarily an operator such as \popp\ would not be primary 
due to the appearance of singular terms with lower derivatives 
of $\phi$ in the OPE with the stress tensor. However 
the mode expansion \phg\ for $\phi$ implies
\eqn\ftp{(\p_\sigma)^k \phi(0,\tau)=0, ~~~~{\rm for}~~~k<h,}
so that \popp\ is indeed primary. 

\subsec{A Bound on the Mass}

The analysis of the previous section was relevant for 
small dimensionless Newton's constant $G_N$ and $m^2\ell^2$ and 
$h$ of order one. In string theory or Kaluza-Klein theory 
one encounters arbitrarily large values of $m$ and so one is 
forced to consider the case $h \sim {1 \over G_N}$. 
In this case conformal invariance in a 
semiclassical expansion requires that \popp\ 
is replaced by the dressed operator\foot{The coefficient of 
$\rho$ in the exponent will be altered if there are 
leading semiclassical corrections to the 
dimension of $e^{-h\rho}$. This is indeed the case for 
the pure Liouville theory considered in \refs{\mms,\sei}.}
\eqn\ftp{e^{-h\rho(0, \tau)}(\p_\sigma)^h\phi(0,\tau).}
For $h \sim {1 \over G_N}$, the semiclassical saddle point 
must include the $\rho$ insertion on the boundary. This problem 
was considered in \mms\ ( the bulk problem is considered in \sei) 
where it 
was shown semiclassically that the 
corresponding operator insertions produce kinks or deficit angles 
in the boundary of order $h  G_N$. This follows from the 
$\rho$ equation of motion which has a delta function source at the 
operator insertion.  Constraining the deficit angle to be less than 
$\pi$ leads to maximum value $h_{max}$ of order 
$1 \over G_N$ for allowed operators. Using the relation 
\hmn\ between the weight and the mass gives a bound on the mass
\eqn\mmx{m_{max} \sim  {1 \over \ell G_N}.} 

In order to get a precise value for $m_{max}$ one must specify 
the gravitational dynamics. The values will be numerically 
different for the pure Liouville theory and the 
dilaton-modified theories encountered in string compactification.
For the moment we simply remark that a bound of the form \mmx\ 
is in qualitative agreement with the stringy exclusion principle 
of \msads, and might provide a generalization of the principle 
to non-BPS states.

\subsec{Remarks}
In this subsection we make several remarks on the preceding 
results.

Roughly speaking one might attempt to decompose the 
central charge as
\eqn\gkl{0=c_{total}=c_{gravity}+c_{matter}+c_{ghosts}.}
The asymptotic growth of physical states, which is related to 
black hole entropy, should be determined by 
$c_{matter}$.
Naively $c_{matter}$ could be computed from the two-point 
function of the matter stress tensor at short distances in the bulk. 
One might expect that every field $\phi$ 
is an independent quantum field and contributes to $c_{matter}$ because
the mass term does not affect the two-point function at sufficiently short 
distances.
However this cannot be correct since there can be an infinite 
number of such fields $e.g.$ from Kaluza-Klein 
compactification or massive string modes. A
bound of the form \mmx\ alleviates, but does not seem 
in general to eliminate the problem. 
Presumably there are further quantum relations 
between the fields ($e.g.$ of the kind encountered in \msads) 
which become very significant at short distances. 

An alternate and possibly more efficient method of determining $c_{matter}$ 
is from the two point function of the stress tensor on the boundary. 
The relation \popp\ implies that $\phi$ does not contribute 
to the two-point function of the stress tensor at the boundary if it has 
weight $h>1$. Hence we need only consider zero or negative (mass)$^2$ fields in 
such a computation.  

It is interesting to contrast the preceding \ad\ analysis 
with the \ads\ analysis of 
\nhbh\ and \msads. In that context the states are also described by a 
conformal field theory, but it lives on the boundary of \ads\ \jbmh, 
whereas in the present 
\ad\ context the conformal field theory is in the bulk of
spacetime. The analysis of the scalar wave equation in \msads\ 
paralleled that 
of the previous paragraph. However it employed $\slr_L \times \slr_R$ rather 
than just a single $\slr$. Furthermore the form of the differential
operators used in the \ads\ case 
was not the familiar one of conformal field theory:
the operators acted on a three dimensional space. The action 
reduces to the familiar form only on the 
boundary of \ads. In the \ad\ context the connection with conformal field
theory is more direct. 

\newsec{\ads\ $\rightarrow$ \ad}
IIB string theory on $K3$ (similar statements pertain to $T^4$) 
has black string solutions 
characterized by D-onebrane (D-fivebrane) charge  $Q_1$ ($Q_5$)
and momentum density ${n \over R^2}$. These objects have 
a dual description as a $c=6Q_1Q_5$ conformal field theory 
whose target space is the symmetric product of $Q_1Q_5$ 
copies of $K3$ \ascv. The near-horizon geometry 
of the spacetime solutions is \ads$\times S^3 \times K3$.  
Compactifying the black string on a circle 
to yield a five-dimensional black hole corresponds to $S^1$ 
compactification of 
\ads\ \hyun . In this section we describe the appearance 
of \ad\ in the very-near-horizon limit of this compactification. 
\subsec{ $S^1$ Compactification}
We begin with the near-horizon \ads\ geometry for the extremal
five-dimensional black hole with charges $(Q_1,Q_5,n)$
\eqn\bsm{ds^2 = {\ell^2}T^2 (dx^5+{dt\over R})^2 + 
{U^2\over {\ell^2}}((Rdx^5)^2-dt^2) + {\ell^2}{(dU)^2\over U^2}.}
$t$ here is the asymptotic time coordinate in the full spacetime solution, 
while the normalization of $x^5$ is such that 
\eqn\uvsim{x^5 \sim x^5 + 2\pi,}
and all factors of the asymptotic $S^1$ radius $R$ are explicit in the
metric. We further
define
\eqn\tdf{\eqalign{T&=\sqrt{n \over k},\cr {\ell^4}&=g_6^2 k,
\cr k&=Q_1Q_5 ,}}
where $g_6$ is the six-dimensional string coupling and 
the dimensionless quantity $T$ is related to the temperature $T_L$
of the left-moving modes of the symmetric product conformal 
field theory by $\pi R T =T_L$. The 
classical supergravity
limit is $g_6 \to 0$ with $g_6^2n,~g_6Q_1$ and $g_6Q_5$ 
fixed and large. 

To get from \ads\ to \ad\ we decompose
\eqn\mtr{ds_3^2=ds_2^2 + \ell^2 e^{2\psi} (dx^5)^2 +\ell^2e^{2\psi} A_t^2dt^2 
+2\ell^2e^{2\psi} A_tdtdx^5,}
with
\eqn\rth{\eqalign{e^{2\psi}&=T^2+{R^2U^2 \over {\ell^4}},\cr
                 A_t&={T^2 \over R}e^{-{2\psi}}.}}
This yields
\eqn\dfg{ds_2^2=-{R^2U^4 \over 
{\ell^6}T^2+{\ell^2}R^2U^2}dt^2 +{{\ell^2} \over U^2}dU^2.}

\subsec{The Very-Near-Horizon Region}
The metric \dfg\ reduces to 
the \ad\ metric in the ``very-near horizon'' limit 
$\lambda \equiv {U^2R^2 k \over \ell^4 n}\to 0$, in which the 
$S^1$ radius $\psi$ is constant. Radii of order 
$U_0$ correspond to excitation 
energies $\mu$ above extremality (see section 3.4 below)
of order $\mu \sim {k^2R^3U_0^4 \over n \ell^8}\sim \lambda^2{n  \over
R}$. Hence for $\lambda \to 0$ one is below the black hole mass gap \msu.
The usual excited states are absent in this region. In 
terms of the dual symmetric-product conformal
field theory  this means the number of right-moving excitations obeys
$n_R =0$. The ground states (characterized by $n_L$) remain and 
should be described in this limit.

In addition to the ground state 
there is the possibility of a continuum of arbitrarily 
low-lying modes corresponding to the separation of the black hole 
into separate pieces. These modes are ignored in 
most discussions of black hole dynamics. In terms of the dual conformal
field theory these are excitations of the Coulumb branch.
This is an interesting subject and perhaps ultimately essential 
for a complete understanding of \ad. However at present we do not 
understand the description of these modes in detail  
and simply ignore them for the rest of the
paper.

One may also consider the 
region of $U^2$  much, but not infinitely smaller than 
${\ell^4 n \over R^2k}$, in which case $\lambda$ is small but nonzero and 
the metric is nearly \ad. $AdS_2/CFT_1$ duality 
should yield approximate relations in this region
in the spirit of \jmtwo. 
This corresponds to excitation 
energies $\mu$ much less than 
${n \over R }$, or $n_R \ll n_L$.  This is the regime of interest 
for the purpose of analyzing 
the near-extremal entropy. 

In the very-near-horizon limit \dfg\ reduces to  
\eqn\aadst{\eqalign{ds^2
&=-{{\ell^2} \over (t^+-t^-)^2}
dt^+dt^- ,}}
with 
$t^\pm=t\pm{{\ell^4}T \over 2RU^2} $.   
The length of the internal circle is 
\eqn\erf{\ell e^{\psi}=\ell \sqrt{n \over k}.}
The Kaluza-Klein gauge field strength is
\eqn\fst{F_{+-}=\p_+A_--\p_-A_+=-{1 \over T(t^+-t^-)^2}={2\epsilon_{+-} \over
\ell^2}\sqrt{k \over n}.}
Note that the asymptotic modulus $R$, which is a fixed scalar, 
drops out of the very-near-horizon geometry. 

The very-near-horizon geometry can be described directly 
as a quotient of \ads\ without a limiting procedure. 
It is useful to transform to Poincare coordinates on \ads\ 
given by
\eqn\ncdr{\eqalign{w^+&={R \over 2 T} e^{2 T(x^5+ {t\over R})}
,\cr
w^-&={ (Rx^5-t) }-{\ell^4  T\over R  U^2}, \cr
y&={\ell^2  \over U} e^{ T(x^5+ {t\over R} )}  }}
in which the metric becomes 
\eqn\dfgo{ds_3^2={{\ell^2} \over y^2}(dw^+dw^-+dy^2).}
The identification $x^5 \sim x^5+2\pi $ corresponds to 
\eqn\wrrz{\eqalign{w^+ & \sim e^{4\pi T} w^+, \cr
w^- & \sim w^-+{2 \pi R },
\cr y&\sim e^{2\pi T } y
.}}
The very-near-horizon limit can be reached by the 
diagonal $SL(2,R)_L\times SL(2,R)_R$ transformation
\eqn\nhf{\eqalign{w^\pm &\to {w^\pm \over \lambda^\prime},\cr y &\to {y\over \lambda^\prime},}}
for $\lambda^\prime \to 0$, since the horizon is at $y=\infty$. 
This transformation takes $R \to {\lambda^\prime R }$, so the
limit is equivalent to 
$R \to 0$. In this case \wrrz\ is a purely $SL(2,R)_L$ 
transformation, and contains no $SL(2,R)_R$ action.
Hence we recover the construction of \dlas\ of the 
near-horizon \ad\ geometry as an $SL(2,R)_L$ quotient of 
\ads. As we shall see later, this leads to a somewhat subtle 
relation between the unbroken $SL(2,R)_R$ isometries of 
\ads\ and the $SL(2,R)$ isometries of \ad.

The two-dimensional \ad\ geometry resulting from the quotient 
\wrrz\ with $R$ set to zero is most simply
expressed in terms of\foot{Alternately, $u=x^5+(t/R)+({\ln}R/2T),
~~t^\pm=Rx^5-t\pm (\ell^4T/RU^2)$.} 
\eqn\wdh{\eqalign{w^-&=t^-,\cr
                  w^+&={1 \over 2T}e^{2Tu},\cr
                  y&=\sqrt{t^+-t^- \over 2T}e^{Tu}.}}
The inverse transformation is 
\eqn\wdh{\eqalign{t^-&=w^-,\cr 
                  t^+&={y^2 \over w^+}+w^-,\cr
                  u&={1 \over 2T}\ln 2Tw^+.}}
One then has
\eqn\glo{\eqalign{
{ds_3^2}&={{\ell^2}(dt^+-dt^-)^2 \over 4(t^+-t^-)^2}+{{\ell^2}Tdu(dt^++dt^-) \over (t^+-t^-)}
+{\ell^2}T^2du^2.}}
Imposing the null identification
\eqn\dtm{u \sim u+2\pi ,}
and using \mtr\ gives
\eqn\djk{\eqalign{e^{2\psi}&=T^2,\cr A_\pm&={1 \over 2T(t^+-t^-)},\cr
ds_2^2&=-{{\ell^2}dt^+dt^- \over (t^+-t^-)^2}.}}

The coordinates $t^\pm$ cover only the region outside the 
horizon, as illustrated. 
Coordinates which cover all of (the universal cover of) \ad\
are\foot{Alternately, for $\tau=u^++u^-, ~~~{\rm sech} \rho  = \sin (u^+-u^-),
~~~4ds^2=-\ell^2\cosh^2\rho d\tau^2+{\ell^2}d\rho^2.$ }
$t^\pm= \tan u^\pm$. The metric is 
\eqn\bgj{ds^2=-{\ell^2 du^+du^- \over \sin^2 (u^+-u^-)}.}
$\sigma=(u^+-u^-)$ runs from $0$ to $\pi$  while $\tau=u^++u^-$ 
is unrestricted.  The horizons are at 
$u^\pm=\pm{\pi \over 2}$. 

Note that \ad\ has two timelike boundaries $\sigma =0,\pi$, 
one inside and one outside the horizon. 
The electric field strength \fst\ can be viewed as arising 
from charges  $ {k \ell^2 \over 2}e^{3\psi}\epsilon^{+-}F_{+-}=\pm {n}$ 
fixed at the  boundaries.  

\subsec{The Dual Description}
String theory on \ads\ has a dual description as a 
$c=6k$ (deformed) symmetric 
product conformal field theory. This theory lives on the cylindrical 
boundary of \ads\ 
with $y=0$. This boundary has null coordinates $w^\pm$. The 
action of the identification \wrrz\ in the 
very near horizon limit for which $R\to 0$ is 
simply a shift of $\ln w^+$, with no action on 
$w^-$. Hence the dual representation of the very-near 
horizon string theory is a DLCQ 
conformal field theory. This suggests a connection 
with the matrix model \refs{\bfss, \dlcq}. 

We note that the one dimension of the $D=1$ conformal field 
theory dual to \ad\ string theory is a null 
($w^-$) coordinate. This provides a natural framework for the 
action of the Virasoro algebra, whose existence was argued 
for on general grounds in section 2. On the other hand the 
general arguments of \refs{\jm,\withol}
suggest a formulation of the dual theory as a conformal quantum mechanics 
on the two timelike boundaries of \ad. It would be interesting to understand 
this formulation in detail. 

It is intriguing to observe that the general 
$AdS_{D}\leftrightarrow CFT_{D-1}$ duality
boils down in this $D=2$ context to a duality between 
two conformal field theories,
one with $c=0$ on the strip and another with $c=6k$ on the 
DLCQ cylinder. Perhaps this 
$D=2$ duality can be understood in more conventional field theoretic terms.

\subsec{Two-Dimensional Effective Action}

The gravity sector in \ads\ is governed by the action
\eqn\str{S_3= { k \over 4\pi \ell} \int d^3 x\sqrt{-g}(R+{2\over
\ell^2}).}
Additional fields suppressed here 
include the $SU(2)_L\times SU(2)_R$ level $k$ 
Chern-Simon gauge 
fields. 
\str\ reduces to a two-dimensional metric, scalar and $U(1)$ gauge field 
governed by\foot{In conformal gauge, $S_2={k \over 2}\int d^2x 
(4 e^\psi \p_-\p_+\rho+{1 \over \ell^2}e^{\psi+
2\rho}+\ell^2 e^{3\psi-2\rho}F_{+-}^2)$. For fixed charge the last term is 
${k \over 2}\int d^2x e^{-3\psi+2\rho}({ n \over \ell k})^2)$ .}
\eqn\strt{S_2={  k \over 2 } \int d^2 x\sqrt{-g}\left(e^{\psi}(R
+{2\over
\ell^2} )-{\ell^2\over 4}e^{3\psi}F^2\right).}
This action has a solution with $R=-{8 \over \ell^2}$ and constant $
\psi$ and $F$ given by
\erf\ - \fst, corresponding to the extremal black hole. Additionally there 
are near extremal solutions given by
\eqn\dsf{\eqalign{ds_2^2&=-{U^2(U^2-U_0^2)\over \ell^2(U^2+U_0^2\sinh^2\sigma)
}dt^2+{\ell^2 \over U^2-U_0^2}dU^2,\cr
e^{2\psi}&={U^2+U_0^2\sinh^2\sigma \over \ell^4},\cr
A_t&={U_0^2\sinh 2 \sigma \over 2 (U^2+U_0^2\sinh^2\sigma )},}}
with $U_0$ and $\sigma $ defined by 
\eqn\wga{\eqalign{n&={kU_0^2\sinh 2\sigma \over 2\ell^4},\cr
\mu&={kU_0^2e^{- 2\sigma} \over 2\ell^4}.}}
$n$ is related to the momentum around the $S^1$ 
while $\mu$ is the energy 
above extremality.

Note that for large $U$ and nonzero $\mu$, 
the radius $\psi$ blows up. This means that  asymptotically one 
effectively returns to \ads.\foot{$\psi$ is an $m^2\ell^2=8$
fixed scalar which according to \hmn\ corresponds to an irrelevant $h=2$
perturbation of the \ad\ conformal field theory. Hence exciting this field 
corresponds to changing the \ad\ boundary conditions. }
Hence the near extremal solution can be understood in the \ad\ 
context only for sufficiently small $U$ and $U_0$. This requires a
small energy $\mu$. 
To leading non-trivial order for such small 
$\mu $ and fixed $n$ the entropy is 
\eqn\dfl{S_{BH}={2 \pi k } e^{\psi (U_0)}=2\pi\sqrt{kn}+\pi \sqrt{2k\mu}.}

\newsec{Symmetries}

In this section we consider the symmetries of \ad\ string theory 
and its relation to those of \ads. 

\subsec{The Unbroken $\slr$}

The quotient \wrrz\ in the very-near-horizon limit 
$R\to 0$ leaves intact the full $\slr_R$ 
isometry group of \ads. One expects this group 
is related to the $\slr$ isometry group of \ad. 
Using the coordinate transformations \wdh\ and the \ads\ isometries one 
finds that the \ads\ and \ad\ isometries are related 
by
\eqn\vrplane{\eqalign{
 H_{-1} = &  i {\p \over \p w^-}  =i({\p \over \p t^+}+{\p \over \p t^-}),
\cr
 H_{0}  = & i ( w^- {\p \over \p w^-} + {1\over 2} y{\p \over \partial y} )
=i(t^+ {\p \over \p t^+} +t^-{\p \over \p t^-} )
\cr
 H_1 = & i ( (w^-)^2{\p \over \p w^-}   +  w^- y {\p \over \partial y}  
- y^2  {\p \over \p w^+} ),
=i(t^{+2}{\p \over\p t^+}   +t^{-2}{\p \over\p t^-} -{t^+-t^- \over 2T}{\p 
\over \p u}).}}
We see that the $ H_{-1}$ and $H_{0}$ generators are identified for 
\ad\ and \ads, but there is a shift in $H_1$. Since $u$ is a 
coordinate on the compactified $S^1$, the extra term can be 
interpreted as a gauge transformations. Hence the $\slr_R$ 
\ads\ isometries reduce to the $\slr$ \ad\ isometry plus a gauge 
transformation. 

Given such a relation between the \ad\ and \ads\ isometry groups 
one naturally expects a relation between the full conformal groups. 
To find this we note that the \ad\ action in \vrplane\ can be written as 
\eqn\erf{H_j=i\bigl( \epsilon_j^+\p_+-{ \p_+ \epsilon_j^+ \over 4T}\p_u+
\epsilon_j^-\p_-+{\p_- \epsilon_j^-  \over 4T}\p_u\bigr),}
with
\eqn\dsc{\epsilon_j^\pm=(t^\pm)^{j+1},}
for $j=0, \pm 1$.  For general integral values of $-\infty <j<\infty$ 
one easily finds 
that the $H_k$ obey the classical Lie bracket relation
\eqn\lbr{ \{ H_j ,H_k \}_{L.B.}=i(j-k)H_{j+k}.} 
One also finds for all values of $j$ using the coordinate 
transformation 
\wdh\ that the action of 
$H_k$ near the boundary $y=0$ of \ads\ is  
\eqn\bvl{ H_k =  i \bigl( \zeta^-_k{\p \over \p w^-}+\p_-\zeta^-_k{y \over 2}
{\p \over \p y}- \p_-^2\zeta^-_k {y^2 \over 2}{\p \over \p w^+}\bigr)\bigl(1+ 
{\cal O}(y^2)\bigr),}
where $\zeta^-_k=(w^-)^{k+1}$. This expression - including the subleading 
terms indicated - is the same as that given in \jbmh\ for diffeomorphisms 
preserving the \ads\ boundary conditions. Hence the $H_k$ generate 
the right-moving \ads\ conformal transformations.

The precise form for all values of $y$ 
of the generators \vrplane\ 
of the \slr\ subgroup is fixed by demanding that they are 
Killing vectors of the \ads\ metric. The other generators 
are identified only by their action at infinity and in general 
have no preferred form in the interior. \erf\ for general $j$ is a
convenient 
choice for our purposes.

Let us now assume that we can define a quantum theory on \ad\ with 
operators $T$ and $U$ corresponding to the full $c=0$ stress tensor and
the appropriately normalized 
conformal current associated to the Kaluza-Klein $U(1)$ 
gauge symmetry. 
Then the diffeomorphisms \erf\ are generated by moments of the twisted stress
tensor
\eqn\fkl{\tilde T_{\pm \pm}=T_{\pm \pm} \mp \p_\pm U_\pm.}
Twisted Virasoro generators 
$\tilde L_n$ can be defined in terms of $\tilde T$. 
This yields the general relation 
between the right Virasoro generators $L_n^{(3)}$ of \ads\ and those of 
\ad :
\eqn\egl{L_n^{(3)}=\tilde L_n=L_n+(n+1)U_n.}
Energy conservation $T_{--}=T_{++}$ as well as 
$\tilde T_{--}= \tilde T_{++}$ on the boundary
requires that the current $U$ obeys Dirichlet boundary conditions:
\eqn\dbc{U_+(\sigma )=-U_-(\sigma )~~~~{\rm for} ~~ \sigma =0,\pi,}
with respect to which the modes of $U$ in \egl\ are defined.
These boundary conditions correspond to fixed 
total charge or fixed electric flux 
\fst.\foot{The charge density $U_t$ does not include 
the fixed boundary charges $\pm {n}$. }
The level of the $U(1)$ current algebra can be deduced from \egl\ 
and the assumption 
that the \ads\ central charge is $c=6k$ while the \ad\ central 
charge is $c=0$. One finds 
\eqn\ucc{[U_m,U_n]=-{km
                  \over 2}\delta_{m+n}.}
It would be of interest to compute this level directly.

The assumption that a suitable quantum theory can be defined on \ad\ 
is perhaps not as benign as it sounds. In order to do so one must 
specify boundary conditions on all the fields, and perhaps also 
understand the ground state entropy in \dfl. The problem is analogous to that  
of finding the Hilbert space in DLCQ, and similar subtleties 
may be encountered. A further subtlety is the possibility of 
singleton degrees of freedom which are confined to the boundary of the strip
but could interact with the bulk degrees of freedom. 
We note that the \ad\ Hilbert space is not the 
same as the \ads\ Hilbert space: they should 
agree only after the Virasoro constraints are imposed on the former. 
Indeed if the two-dimensional $L_n$s are translated back to \ads\ they 
do not generate diffeomorphisms of the form \bvl\ - the last term is missing -
and hence create states which fail to obey the \ads\ boundary conditions of 
\jbmh\ by a coordinate transformation. Hence the 
unconstrained \ad\ Hilbert space 
is larger than \ads\ by spurious states, which are hopefully 
eliminated by constraints. 

In addition to the $U(1)$ gauge symmetry arising from the $S^1$ isometry 
there is an $SU(2)_L\times SU(2)_R$ gauge symmmetry from the 
isometries of $S^3$.
This appears to lead to the large $N=4$ $A_\gamma$ algebra \stp. Since $c=0$ 
it is the degenerate case $\gamma=0$. This algebra is known \stp\ to contain 
the standard $N=4$ algebra as a subalgebra. The relation between the Virasoro
generators is exactly \egl. 

The twisting \egl\ of the stress tensor has also appeared as a central 
formula in the study of topological field theory \wtop.  
In that context one begins with a conventional supersymmetric 
conformal field theory with a 
unitary Hilbert space. This corresponds to the \ads\ conformal 
field theory in the present 
discussion. The stress tensor is then twisted to shift the central charge 
to zero exactly as in \egl. Once the central charge is zero, the theory 
can be reinterpreted as a theory of gravity. Since the partition function for 
gravity is independent of a choice of background metric this is a topological field theory.

\subsec{Charged Kaluza-Klein Modes}

\ads\ string theory has a tower of scalar fields in short 
BPS multiplets of various masses 
as described in \refs{\msads, \sez, \dbo}.  For every scalar field of mass 
$m$ on \ads\ there is a corresponding scalar field of mass $m$ on \ad\ 
obtained from the constant mode on $S^1$. For each of these fields 
there is a highest weight state with weight given by \hmn\ as $h=\half(1+\sqrt{1+
m^2\ell^2})$, and a corresponding primary operator in the boundary conformal 
field theory. 

The full Kaluza-Klein tower of fields $\phi_{q}$ on \ad\ arises from $S^1$ modes 
$e^{iqu}$, where $q$ is an integer. These states have charge $q$ and mass 
\eqn\meq{m_q^2=m^2+{q^2\over \ell^2 T^2},}
where $m$ is the \ads\ mass of the scalar field. 
We wish to find the weight of the highest weight state created by $\phi_{q}$, following the 
$q=0$ analysis of section 2.
The $\phi_q$ wave equation is 
\eqn\pkl{{\cal D}^2\phi_{q}=m_q^2\phi_{q}.}
Using ${\cal D}=\nabla-iqA$, the conformal gauge metric \dfh\ on the 
strip, equation \meq\ and the gauge 
\eqn\aeq{A_-=A_+={\cos (u^+-u^-)\over 2T \sin (u^+-u^-) },}
\pkl\ can be rewritten
\eqn\ghj{\bigl( -4\sin^2(u^+-u^-)\p_+\p_- +{iq \over T}\sin 2(u^+-u^-)(\p_++\p_-)
-{q^2 \over T^2}\sin^2(u^+-u^-) \bigr)\phi_{q}=m^2\phi_{q}.}
In terms of the twisted Virasoro generators 
on the strip
\eqn\tvr{\tilde L_n={i \over 2}(e^{2niu^+}\p_++e^{2niu^-}\p_-)
+{iqn \over4 T}(e^{2niu^+}-e^{2niu^-}).}
\ghj\ has the simple form
\eqn\dth{ -4(-\tilde L_0(\tilde L_0-1)+\tilde L_{-1}\tilde L_1)
\phi_{q}=m^2\ell^2\phi_{q} .}
It follows that the highest weight solutions obeying $\tilde L_1\phi_{h,q}=0$
have a weight $h$ which is independent of $q$ and given by \hmn. 
The wave function is 
\eqn\wfc{\phi_{h,q}=(e^{-2iu^+}-e^{-2iu^-})^he^{-{q \over2 T}(u^+-u^-)}.}
We note that the boundary operators 
$(\p_\sigma)^k\phi_q(0,\tau)$ vanish for $k<h$ independently of $q$. 
To summarize, for every field on \ads\ of mass $m$ and associated weight $h$,
there is an tower of perturbative\foot{A non-perturbative exclusion principle 
could eliminate fields with large $q$.} fields labeled by the Kaluza-Klein 
charge $q$. These have $q$ dependent masses \meq\ but $q$-independent weights 
\hmn. In particular all the charged Kaluza-Klein modes of the $m^2=0$ scalars 
on \ads\ are $h=1$ scalars on \ad. These will all have non-vanishing 
boundary stress tensor and contribute to boundary computation of 
the central charge.

\smallskip
\centerline{\bf Acknowledgements}
   I would like to thank T. Banks, R. Gopakumar, J.Harvey, J. Maldacena, 
E. Martinec, G. Moore, 
S. Shenker, 
C. Vafa and A. Volovich for useful conversations. I am grateful to the Institute 
for Theoretical Physics at Santa Barbara and the Aspen Center for 
physics for hospitality while parts of this 
work were carried out. The work was supported in part by DOE grant 
DE-FG02-91ER40654.
\listrefs
\end